\documentclass[letterpaper,english,twocolumn]{revtex4}
\usepackage[T1]{fontenc}
\usepackage[latin9]{inputenc}
\usepackage{graphicx}
\usepackage{epstopdf}



\usepackage{babel}

\begin{document}

\title{Small-World Bonds and Patchy Percolation on the Hanoi Network}
\author{Stefan Boettcher}
\homepage{http://www.physics.emory.edu/faculty/boettcher/}
\author{Jessica L. Cook}
\affiliation{Emory University, Dept.~of Physics, Atlanta, GA 30322}
\author{Robert M. Ziff}
\email{rziff@umich.edu}
\affiliation{Center for the Study of Complex Systems and Dept.~of Chemical Engineering,
University of Michigan, Ann Arbor, MI 48109-2136 USA}

\begin{abstract}
The bond-percolation properties of the Hanoi networks are analyzed
with the renormalization group. Unlike scale-free networks, they are
meant to provide an analytically tractable interpolation between finite
dimensional, lattice-based models and their mean-field limits. In
percolation, the hierarchical small-world bonds in the Hanoi networks
impose a new form of order by uniting otherwise disconnected, local
clusters. This {}``patchy'' order results in merely a finite probability
to obtain a spanning cluster for certain ranges of the bond probability,
unlike the usual 0-1 transition found on ordinary lattices. The various
networks studied here exhibit a range of phase behaviors, depending
on the prevalence of those long-range bonds. Fixed points in general
exhibit non-universal behavior. \break
\end{abstract}
\maketitle

\section{Introduction\label{sec:Introduction}}
Percolation---the formation of a large connected component---is a
geometric property in the arrangement of a many-body system that can
strongly impact its physical behavior \citep{Stauffer94,Sahimi94}.  Typically,
to observe any form of emergent, collective phenomena, an extensive
fraction of the degrees of freedom must be sufficiently
interconnected. Examples abound, from transport in
amorphous materials \citep{Isichenko92} or the onset of plasticity in complex fluids \citep{Degennes76} to
the spreading of rumors or disease in social networks. In particular,
while percolation on regular and fractal lattice geometries
\citep{Stauffer94} and ordinary random graphs \citep{ER,Bollobas} is a
well-developed industry, its properties and impact of transport on the
many conceivable forms of engineered networks (random, scale-free,
etc) are just beginning to be explored \citep{Barabsi03,NewmanSIAM03,Boccaletti06}.

Here, we study the percolation properties of the recently introduced
set of Hanoi networks \citep{SWPRL,SWN,SWlong}. These networks mimic
the behavior of small world systems without the usual disorder
inherent in the construction of such networks. Instead, they attain
these properties in a recursive, hierarchical manner that lends itself
to exact renormalization \citep{Plischke94}. These networks do not
possess a scale-free degree distribution
\citep{Barabasi99,Andrade05,Zhang07}; they are, like the original
Watts-Strogatz ``small worlds'' \citep{Watts98,Newman99}, of regular
degree or have an exponential degree distribution. Yet, in turn the
Hanoi networks have a more {}``physically'' desirable geometry.  In
particular, they have the potential to provide an analytically
tractable interpolation between finite dimensional, lattice-based
models and their mean-field limits (although no such interpolation
will be considered here \citep{SWN}.) Scale-free networks geared
towards social phenomena or other lattices often used to analyze
physical systems, such as the hierarchical lattice derived from the
Migdal-Kadanoff bond-moving scheme
\citep{Migdal76,Kadanoff76,Berker79}, do not have a physically
relevant mean-field behavior.

Many of the phenomena of interest in statistical physics, such as
critical behavior, are fundamentally related to their percolation
properties. We find that the renormalization group (RG) applied to the
Hanoi networks behaves very differently from the traditional approach
but similar to a low-dimensional hierarchical lattice with small-world
bonds that was recently introduced \citep{Hinczewski06,Berker09}.
This is surprising, as this behavior has usually been associated with
scale-free graphs. Due to the hierarchical nature of the small-world
bonds, the RG possesses unrenormalized parameters, which change the
character of the RG-flow in unusual ways. In the exactly obtained
phase diagrams, stable and unstable fixed points are drawn out on
lines that often merge in branch points, generally making the scaling
near fixed points parameter-dependent and, hence, non-universal. As
the flow can get attracted onto a stable line, entirely new, mixed
ordered-disordered phases appear in what we call {}``patchy'' order:
Otherwise isolated patches of localized clusters at low levels of the
hierarchy get reconnected globally at higher levels with finite
probability (depending on said parameters) to attain long-range
order. In the percolation problems discussed here, this manifests
itself in the fact that for a fixed bond-probability the infinite
network may percolate with a finite probability between zero and
unity. (In conventional percolation, that probability has an instant
0-1 transition at the critical bond density $p_{c}$, although the size
of the spanning cluster varies continuously, see Fig.~17 in
Ref.~\citep{Stauffer94}).

Our paper is structured as follows: In the next section, we review the
construction of the Hanoi networks, followed by a pedagogical
discussion of the renormalization group applied to percolation on
hierarchical lattices with small-world bonds in
Sec.~\ref{sec:MKperco}.  Then, in Sec.~\ref{sec:HNperco} we study the
percolation properties of the Hanoi networks in some depth, and we
conclude in Sec.~\ref{sec:Conclusions}.

\begin{figure}
\includegraphics[clip,scale=0.3]{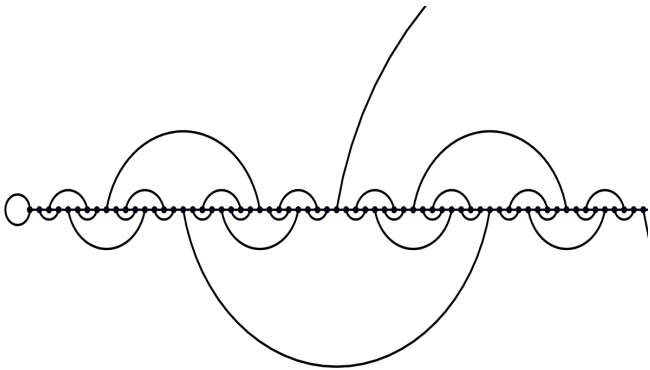}
\caption{\label{fig:3hanoi}Depiction of the 3-regular network HN3 on a semi-infinite
line.  The entire graph becomes 3-regular with a self-loop at $n=0$.
Note that HN3 is planar.}
\end{figure}

\section{Geometry of the Hanoi Networks\label{sec:Graph-Structure}}
Each of the Hanoi network possesses a simple geometric backbone, a
one-dimensional line of sites $n$, either infinite 
$\left(-2^{k}\leq n\leq2^{k},\, k\to\infty\right)$,
semi-infinite $\left(0\leq n\leq2^{k},\, k\to\infty\right)$, or closed
into a ring of $N=2^{k}$ sites. Most importantly, all sites are connected
to their nearest neighbors, ensuring the existence of the $1d$-backbone.
To generate the small-world hierarchy in these networks, consider
parameterizing any integer $n$ (except for zero) \emph{uniquely}
in terms of two integers $(i,j)$,  via
\begin{eqnarray}
n & = & 2^{i}\left(2j+1\right).
\label{eq:numbering}
\end{eqnarray}
Here, $i\geq0$ denotes the level in the hierarchy whereas $j$ labels
consecutive sites within each hierarchy. For instance, $i=0$ refers
to all odd integers, $i=1$ to all integers once divisible by 2 (i.
e., 2, 6, 10,...), and so on. In these networks, aside from the backbone,
each site is also connected with (one or both) of its nearest neighbors
\emph{within} the hierarchy. For example, we obtain a 3-regular network
HN3 (best done on a semi-infinite line) by connecting first the backbone,
then 1 to 3, 5 to 7, 9 to 11, etc, for $i=0$, next 2 to 6, 10 to
14, etc, for $i=1$, and 4 to 12, 20 to 28, etc, for $i=2$, and so
on, as depicted in Fig.~\ref{fig:3hanoi}. (The corresponding 4-regular
network HN4 \citep{SWPRL}, which we will not study here, is obtained
in the same manner but connecting to \emph{both} nearest neighbors
in the hierarchy, i.~e., 1 to 3, 3 to 5, 5 to 7, etc, for $i=0$,
2 to 6, 6 to 10, etc, for $i=1$, and so forth.)

Previously \citep{SWPRL}, it was found that the average chemical path
between sites on HN3 scales with the distance $l$ as 
\begin{equation}
d^{HN3}\sim\sqrt{l}
\label{eq:3dia}
\end{equation}
along the backbone. In some ways, this property
is reminiscent of a square-lattice consisting of $N$ lattice sites,
whose diameter (=diagonal) is also $\sim\sqrt{N}$.

\begin{figure}
\includegraphics[bb=0bp 10bp 600bp 700bp,clip,angle=-90,scale=0.3]{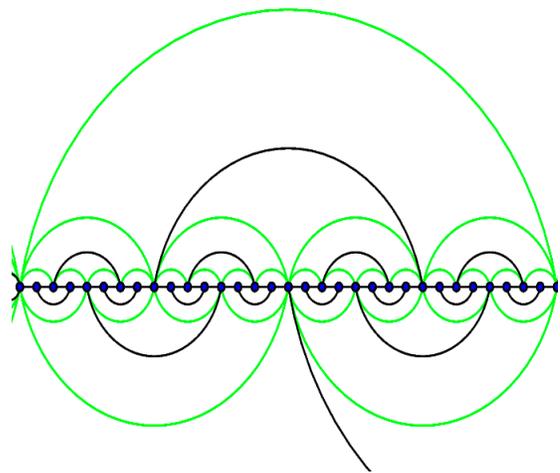}
\caption{\label{fig:5hanoi}Depiction of the planar network HN5, a comprised
of HN3 (black lines) with the addition of further long-range bonds
(green-shaded lines). There is no distinction made between black and
shaded lines in our studies here.}
\end{figure}

While HN3 (and HN4 \citep{SWPRL}) are of a fixed, finite degree, we
introduce here convenient generalizations of HN3 that lead to new,
revealing insights into small-world phenomena. First, we can extend
HN3 in the following manner to obtain a new planar network of average
degree 5, hence called HN5, at the price of a distribution in the
degrees that is exponentially falling. In addition to the bonds in
HN3, in HN5 we also connect all even sites to both of their nearest
neighboring sites that are \emph{within} the same level of the
hierarchy $i\geq1$ in Eq.~(\ref{eq:numbering}). The resulting network
remains planar but now sites have a hierarchy-dependent degree, as
shown in Fig.~\ref{fig:5hanoi}. To obtain the average degree, we
observe that $\frac{1}{2}$ of all sites have degree 3, $\frac{1}{4}$
have degree 5, $\frac{1}{8}$ have degree 7, and so on, leading to an
exponentially falling degree distribution of ${\cal P}\left\{
\alpha=2i+3\right\} \propto2^{-i}$. Then, the total number of bonds
$L$ in the system of size $N=2^{k}$ is
\begin{eqnarray}
2L & =2k-1+ & \sum_{i=0}^{k-1}\left(2i+3\right)2^{k-1-i}=5\,2^{k}-6,
\label{eq:TotalLinksHN5}
\end{eqnarray}
and thus, the average degree is 
\begin{eqnarray}
\left\langle \alpha\right\rangle  & = & \frac{2L}{N}\sim5.
\label{eq:averageDegreeHN5}
\end{eqnarray}

In HN5, the end-to-end distance is trivially 1, see Fig.~\ref{fig:5hanoi}.
Therefore, we define as the diameter the largest of the shortest paths
possible between any two sites, which are typically odd-index sites
furthest away from long-distance bonds. For the $N=32$ site network
depicted in Fig.~\ref{fig:5hanoi}, for instance, that diameter is
5, measured between site 3 and 19 (starting with $n=0$ is the left-most
site), although there are many other such pairs. It is easy to show
recursively that this diameter grows strictly as
\begin{eqnarray}
d^{HN5} & = & 2\left\lfloor k/2\right\rfloor +1\sim\log_{2}N.
\label{eq:5dia}
\end{eqnarray}
We have checked numerically that the \emph{average} shortest path
between any two sites appears to increase logarithmically with system
size $N$ as well.

\begin{figure}
\includegraphics[bb=0bp 0bp 760bp 450bp,clip,scale=0.3]{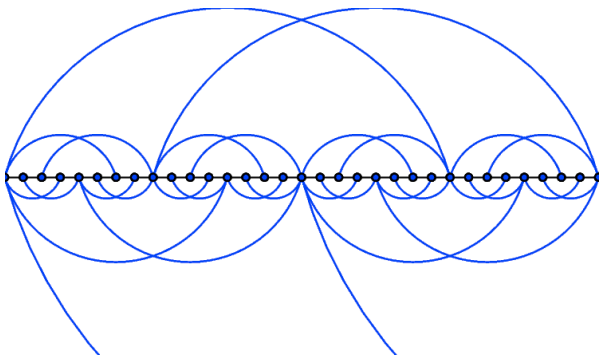}
\caption{\label{fig:hanoi_nonplanar}Depiction of the non-planar Hanoi network
HN-NP. Again, starting from a $1d$-backbone (black lines), a set
of long-range bonds (blue-shaded lines) is added that break planarity
but maintain the hierarchical pattern set out in Eq.~(\ref{eq:numbering}).
The RG on this network remains exact. }
\end{figure}

The networks HN3 and HN5 have the convenient but (from a mean-field
perspective) unrealistic restriction of being planar. In fact, with
a minor extension of the definition, it is easy to also design Hanoi
networks that are both non-planar \emph{and} fully renormalizable.
The simplest such network, which we dub HN-NP here, is depicted in
Fig.~\ref{fig:hanoi_nonplanar}. (Extensions will be introduced
in Ref.~\citep{HanoiIsing09}.)

To obtain the average degree, we observe that $\frac{1}{2}+\frac{1}{4}$ of all sites
have degree 3, $\frac{1}{8}$ has degree 5, $\frac{1}{16}$ has degree 7, and so on,
leading to an exponentially falling degree distribution, as for HN5.
The total number of bonds $L$ in the system of size $N=2^{k}$ is
\begin{eqnarray}
2L & = & 3\,2^{k-1}+\sum_{i=2}^{k-1}\left(2i-1\right)2^{k-i}+2k+(2k-2)
\nonumber\\
 & = & 4\,2^{k}-4,\label{eq:TotalLinksHN-NP}
\end{eqnarray}
and thus, the average degree is 
\begin{eqnarray}
\left\langle \alpha\right\rangle  & = & \frac{2L}{N}\sim4.
\label{eq:averageDegreeHN-NP}
\end{eqnarray}
Here, too, it is easy to see that the shortest paths between sites
increases logarithmically with system size $N$.

\begin{figure}
\includegraphics[bb=0bp 300bp 430bp 710bp,clip,scale=0.3]{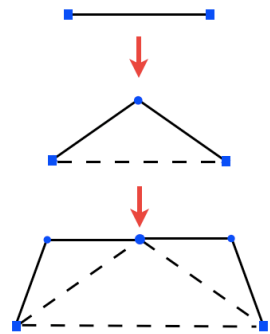}
\caption{\label{fig:MK1}Pattern for the recursive generation of a hierarchical
lattice. In zeroth generation (top), the lattice consists of a single
bond between the two {}``boundary'' sites (squares). In successive
generations each bond of that type (solid line) is replaced by a new
type of bond (dashed line) in parallel with a sequence of two of the
original bonds (solid lines), see middle and bottom for generation
one and two, resp.}
\end{figure}

\section{RG for Percolation\label{sec:MKperco}}
The principles of the renormalization group (RG) for percolation as
it will be use in this paper are easiest to explain for a simple hierarchical
lattice \citep{Berker79,Hinczewski06}. (Percolation on these hierarchical
lattices has recently also been discussed in Ref.~\citep{Berker09}.)
In this recursively constructed lattice, bonds from a preceding generation
$g$ are built up in the next generation $g+1$ according to a fixed
pattern, until the complete network is constructed at a generation
$g=k\to\infty$. Applying real-space renormalization to the complete
network in effect reverses the recursive built-up of the network.
The complete network with the {}``bare'' operators obtained in the
$g=k$-th generation provides the initial conditions, or the $n=0$-th
step in the RG, which we evolve to $n\to\infty$ (i.~e., $k\to\infty$)
and study the fixed points of this recursion and their stability. 

An example of this is shown in Fig.~\ref{fig:MK1} for the first three
generations ($g=0,1,2$) of a hierarchical lattice that corresponds
to a one-dimensional line (solid bonds) with a hierarchy of small-world
bonds (dashed lines) \citep{Hinczewski06}. It is fruitful to regard
both types of bonds as distinct at this stage, although in any complete
network one may be interested mostly in the case where all bonds are
indistinguishable and exist with equal probability in the context
of percolation, say. 

\begin{figure}
\includegraphics[bb=0bp 230bp 610bp 750bp,clip,scale=0.3]{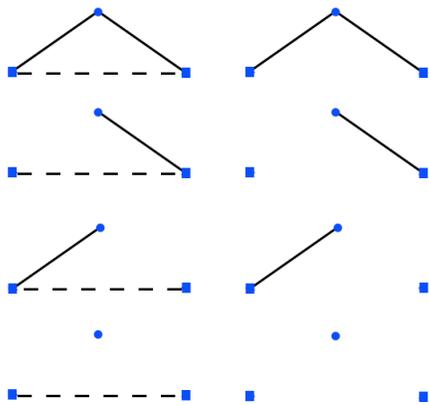}
\caption{\label{fig:MK1perco}All possible combinations of reducing bonds from
the $n$-th RG step to form a new bond at step $n+1$, reversing the
pattern set in Fig.~\ref{fig:MK1}. All graphs on the left percolate
(i.~e., connect end-to-end) by merit of the dashed bond alone, while
on the right only the top graph provides such a connection. If $q_{n}$
is the probability for a solid and $p$ for a dashed bond, the weights
for the graphs on the left are $pq_{n}^{2}$, $pq_{n}\left(1-q_{n}\right)$,
$pq_{n}\left(1-q_{n}\right)$, and $pq_{n}\left(1-q_{n}\right)^{2}$
and on the right are $(1-p)q_{n}^{2}$, $(1-p)q_{n}\left(1-q_{n}\right)$,
$(1-p)q_{n}\left(1-q_{n}\right)$, and $(1-p)q_{n}\left(1-q_{n}\right)^{2}$
from top to bottom.}
\end{figure}

\begin{figure}[b!]
\includegraphics[clip,scale=0.3]{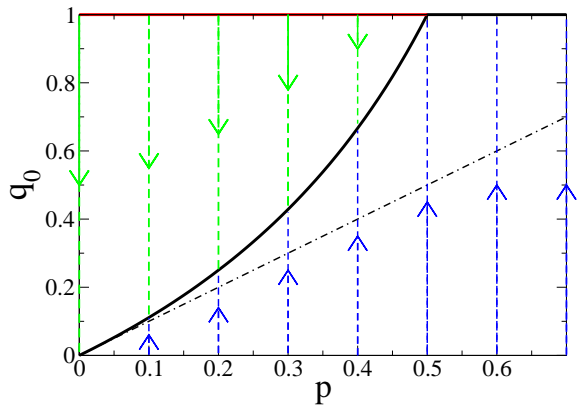}
\caption{\label{fig:MK1phase}Phase diagram for the RG flow of the simple hierarchichal lattice due to Eq.~(\ref{eq:RGrecurMK1}).
For any given $p$, the flow for $q_{n}$ evolves along the flow trajectories
starting from a chosen initial value $q_{0}$. Except for points $\left(p<\frac{1}{2},q_{0}=1\right)$
on the (thick red) line of unstable fixed points, all points $\left(p,q_{0}<1\right)$
flow towards the (thick black) line of stable fixed points from Eq.
(\ref{eq:RG_MK1_fp}) or $q^{*}=1$ for $p>\frac{1}{2}$. This is
especially true for a homogeneous choice of bond probabilities in
a network, $q_{0}=p$ along the (dash-dotted) diagonal line.}
\end{figure}

In Fig.~\ref{fig:MK1perco}, we list all 8 graph-lets consisting of
those three bonds. Summing the weights of those 5 graph-lets that
connect end-to-end and, thus, contribute to the percolation probability
$q_{n+1}$, we obtain
\begin{eqnarray}
q_{n+1} & = & p+(1-p)q_{n}^{2}.
\label{eq:RGrecurMK1}
\end{eqnarray}
It should be noted that there is no corresponding recursion for the small-world
bond $p$ (dashed line). It enters the RG anew at each step, bare
and unrenormalized. (As we will see, this feature is a defining characteristic
also of RG in the Hanoi networks, as discussed in Ref.~\citep{SWPRL}.)

It is very simple, or course, to obtain the fixed points of the RG
in Eq.~(\ref{eq:RGrecurMK1}) in the thermodynamic limit $k\to\infty$
with the Ansatz $q_{n+1}=q_{n}=q^{*}$ for $n\to\infty$. Our case
yields the trivial unstable fixed point $q^{*}=1$ and a non-trivial
but \emph{stable} fixed point at 
\begin{eqnarray}
q^{*}(p) & = & \frac{p}{1-p}\qquad\left(p\leq\frac{1}{2}\right).
\label{eq:RG_MK1_fp}
\end{eqnarray}
We have obtained a line of stable fixed points depending on the parameter
$p$. For $p=0$, we obtain the ordinary and trivial percolation problem
of a one-dimensional lattice with a percolation threshold at $q_{c}=1$:
For any initial bond probability $q_{0}<1$, the RG flow from the
recursion in Eq.~(\ref{eq:RGrecurMK1}) evolves away from the unstable
fixed point $q^{*}=1$ to the stable fixed point $q^{*}=0$. The ordered
(percolating) state is attained only for $q_{0}=1$. For all $p\geq0$,
the phase diagram for the RG flow is shown in Fig.~\ref{fig:MK1phase}.
For any $p<\frac{1}{2}$, the usual ordered state remains confined
to the unstable fixed point $q_{0}=q^{*}=1$. For any other starting
value $0\leq q_{0}<1$, the RG flow converges onto the line of non-trivial
stable fixed points given by Eq.~(\ref{eq:RG_MK1_fp}) in which the
usual non-percolating, disordered state is now replace by a \emph{partially
ordered} state, i.~e., even in the thermodynamic limit there is a \emph{finite}
probability to percolate. For $p\geq\frac{1}{2}$, the stable and
unstable fixed points merge and an ordered, percolating state is reached
irrespective of $q_{0}$ (even when it vanishes), merely by the strength
of the small-world bond, which repeatedly, at any generation of the
hierarchical lattice, has a chance $p$ to rectify any non-percolating
sub-lattices from previous generations (see, for instance, the bottom
network in Fig.~\ref{fig:MK1}). In reference to the corresponding
effect we found previously for the Ising ferromagnet on HN5, we call
this partially ordered state {}``patchy'' \citep{HanoiIsing09}:
ordering occurs between the lucky neighborhoods that happen to be
connected by the occasional small-world bond.

We obtain this picture from a local analysis of Eq.~(\ref{eq:RGrecurMK1})
near the fixed point with the Ansatz
\begin{equation}
q_{n}\sim q^{*}-\delta_{n}
\label{eq:AnsatzMK1}
\end{equation}
assuming $\delta_{n}\ll1$. To leading order with $q^{*}=1$
yields
\begin{eqnarray}
\delta_{n+1} & \sim & 2(1-p)\,\delta_{n}
\label{eq:MK1correction}
\end{eqnarray}
with solution
\begin{eqnarray}
\delta_{n} & \sim & \left[2\left(1-p\right)\right]^{n}\delta_{0}.
\label{eq:MK1stability}
\end{eqnarray}
For $2\left(1-p\right)<1$, the corrections $\delta_{n}$ contract
for increasing $n$, leaving the fixed point stable. In contrast,
for $2\left(1-p\right)>1$ the corrections escalate (eventually violating
the assumption) and the fixed point is said to be unstable. As shown
in Fig.~\ref{fig:MK1phase}, the line of unstable fixed points at
$q^{*}=1$ for small $p$ thus disappears at $\bar{p}=\frac{1}{2}$
and is replaced by a stable fixed point at all larger $p$.

Typically, in a complete network, we make no distinction between the
probabilities of any of the bonds in the network, whether they are
small-world or not. Hence, in light of Fig.~\ref{fig:MK1phase}, the
RG flow would initiate with $q_{0}=p$, which corresponds to the flow
trajectories emanating from the diagonal (dash-dotted) line. Consequently,
projecting the flow in the diagram just onto that diagonal, one would
conclude that there is no conventional phase transition. Starting
on the open diagonal at any point the flow proceeds to a larger effective
percolation probability. There is an interesting transition in this
behavior at $\bar{p}=\frac{1}{2}$ between a fully ordered (percolating)
phase above, where this flow always converges to $q^{*}=1$, and a
partially-ordered, patchy phase below, converging to some non-trivial
value of  $q^{*}=q^{*}(p=q_{0})<1$
given by Eq.~(\ref{eq:RG_MK1_fp}). We will see that this situation
resembles very closely that of percolation on HN5. Notice that the
unstable fixed point discussed above, typically itself the focus of
any study in critical phenomena, seems to have become irrelevant to
these considerations, as our initial conditions never cross an unstable
manifold. 

\begin{figure}
\includegraphics[bb=0bp 400bp 430bp 710bp,clip,scale=0.3]{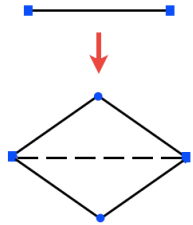}
\caption{\label{fig:MK2}Pattern for the recursive generation of a $2d$ hierarchical
lattice, originating from the Migdal-Kadanoff bond-moving scheme in
a square lattice, with the addition of a small-world bond. Each single,
solid bond between the two sites (squares) of a previous generation
is replaced by a small-world bond (dashed line) in parallel with two
sequences of two of the original bonds (solid lines) at the next generation.}
\end{figure}

\begin{figure}
\includegraphics[clip,scale=0.3]{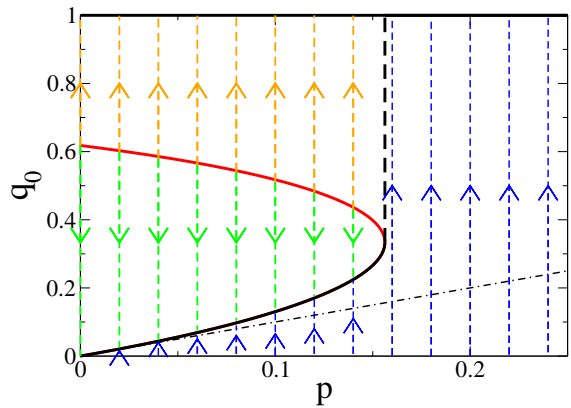}
\caption{\label{fig:MK2phase}Phase diagram for the RG flow of the $2d$ hierarchichal lattice due to Eq.~(\ref{eq:RGrecurMK2}),
using the same notation as in Fig.~\ref{fig:MK1phase}. Here, there
exists an additional phase for $p<\bar{p}=\frac{5}{32}=0.15625$ due
to the (thick red) line of unstable fixed points on some line $q_{0}(p)<1$.
Points above that line flow into the now stable ordered fixed point
$q^{*}=1$. Note that for the homogeneous choice of bond probabilities
in a network, i.~e., $q_{0}=p$ along the (dash-dotted) diagonal line,
we observe a discontinuous behavior at $\bar{p}$: Coming from the
partially ordered phase below, the flow towards the finite percolation
probability $q^{*}<1$ jumps (along the thick dashed line)
towards $q^{*}=1$ in the fully ordered phase above $\bar{p}$.}
\end{figure}

It is, in fact,  just as simple to analyze the corresponding behavior
for the small-world hierarchical network on an underlying two-dimensional
lattice \citep{Hinczewski06}. As shown in Fig.~\ref{fig:MK2}, we
merely need to add a second sequence of two (solid line) bonds to
the previous hierarchical generation of the network in Fig.~\ref{fig:MK1},
resulting in a new graph-let of five bonds. As in Fig.~\ref{fig:MK1perco},
we consider the now $2^{5}=32$ graph-lets regarding the end-to-end
connectivity and sum the weights of those that contribute to percolation.
Analogous to the one-dimensional RG recursion~(\ref{eq:RGrecurMK1})
we find
\begin{eqnarray}
q_{n+1} & = & p+(1-p)q_{n}^{2}\left(2-q_{n}^{2}\right),
\label{eq:RGrecurMK2}
\end{eqnarray}
which is very similar in character: it also depends on a non-renormalized,
bare parameter $p$ for the probabilities of the small-world bond at
every recursion. The same fixed-point analysis as above leads to a
more complex but straightforward algebraic equation for $q^{*}=q^{*}(p)$,
and we plot the respective RG flow in Fig.~\ref{fig:MK2phase}. Without
small-world bonds, at $p=0$, we already observe the emergence of
a nontrivial unstable fixed of the ordinary two-dimensional hierarchical
lattice at $q^{*}=1/\phi=0.618\ldots$, where $\phi=\left(\sqrt{5}+1\right)/2$
is the {}``golden ratio'' \citep{Livio03}. In this case, $q^{*}(p)$
develops a remarkable branch-point singularity within the physical
domain at the critical point $\bar{p}=\frac{5}{32}$, where two branches
of fixed points pinch off into the complex plane and the stable fixed
point attained for all points along the diagonal $q_{0}=p$ jumps
discontinuously from $q^{*}\left(p\to\bar{p}^{-}\right)=\frac{1}{3}$
to $q^{*}\left(p\to\bar{p}^{+}\right)=1$. 

Again, we can verify these findings with a local analysis along the
line of fixed points with the same Ansatz as in Eq.~(\ref{eq:AnsatzMK1}),
where $q^{*}(p)$ is now a solution of Eq.~(\ref{eq:RGrecurMK2}),
see the parabolic line in Fig.~\ref{fig:MK2phase}. We obtain 
\begin{eqnarray}
\delta_{n+1} & \sim &
4\left(1-p\right)q^{*}\left[1-\left(q^{*}\right)^{2}\right]\,\delta_{n},
\label{eq:MK2correction}
\end{eqnarray}
where the simultaneous solution of the fixed point equation
together with the stability condition on the Ansatz,
\begin{eqnarray}
4\left(1-p\right)q^{*}\left[1-\left(q^{*}\right)^{2}\right] & < & 1,
\label{eq:MK2stability}
\end{eqnarray}
defines the branch of stable fixed points, while its violation implies
the unstable branch. Both branches meet when the inequality in Eq.
(\ref{eq:MK2stability}) is saturated, at $\bar{p}=\frac{5}{32}$
and $q^{*}=\frac{1}{3}$. The fixed point analysis for $q^{*}=1$
leads to
\begin{eqnarray}
\delta_{n+1} & \sim & 4\left(1-p\right)\delta_{n}^{2}.
\label{eq:nonlinearMK2}
\end{eqnarray}
This  non-linear relation has the solution
\begin{eqnarray}
\delta_{n} & \sim &
\frac{1}{4\left(1-p\right)}\left[4\left(1-p\right)\delta_{0}\right]^{2^{n}}.
\label{eq:delta_nonlin}
\end{eqnarray}
Deriving from our assumption, it is certainly $\delta_{0}<\frac{1}{4}$
and hence $\delta_{n}$ vanishes and the fixed point proves to be
exceedingly attractive and stable, ever more so for increasing $p$.

\section{RG for Percolation on Hanoi networks\label{sec:HNperco}}
\begin{figure}[t!]
\includegraphics[bb=0bp 0bp 420bp 150bp,clip,scale=0.6]{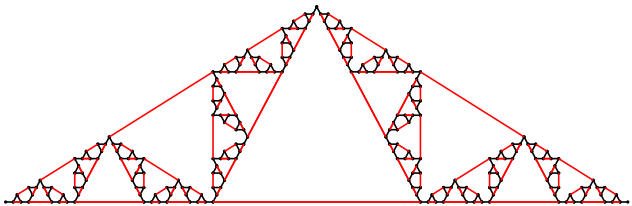}
\caption{\label{fig:3hanoiKoch}Representation of HN3 as a branched Koch curve.
The one-dimensional backbone shown in Fig.~\ref{fig:3hanoi} is marked
in black here; long-range bonds are shaded in red. In this representation,
the diameter in Eq.~(\ref{eq:3dia}) corresponds to the base-line
in the Koch curve. To prevent end-to-end connectivity, only two bonds
right down the symmetry axis need to be cut, independent of 
size. Hence, the network is finitely ramified.}
\end{figure}

In this section, we study percolation on HN3, HN5 and HN-NP with the
renormalization group. It is easy to convince oneself that percolation
on HN3 can only occur at full connectivity $p\to1$, because HN3 is
finitely ramified: To prevent end-to-end connectivity or the emergence
of a giant component only a fixed number of bonds need to be cut at
\emph{any} system size \citep{Stauffer94}. This is most apparent
by plotting HN3 as a branched Koch curve, see Fig.~\ref{fig:3hanoiKoch}.
Yet, the limit $p\to1$ in HN3 itself is interesting, showing logarithmic
finite-size corrections, which appear impossible to resolve numerically.
Moreover, the derivation of the RG recursions for HN3 are almost identical
to that for HN5 and HN-NP, which in turn have a more interesting phase
diagram. 

\subsection{Renormalization Group Studies for HN3\label{sub:RG_HN3}}
\begin{figure}[t!]
\includegraphics[bb=0bp 20bp 369bp 200bp,clip,scale=0.5]{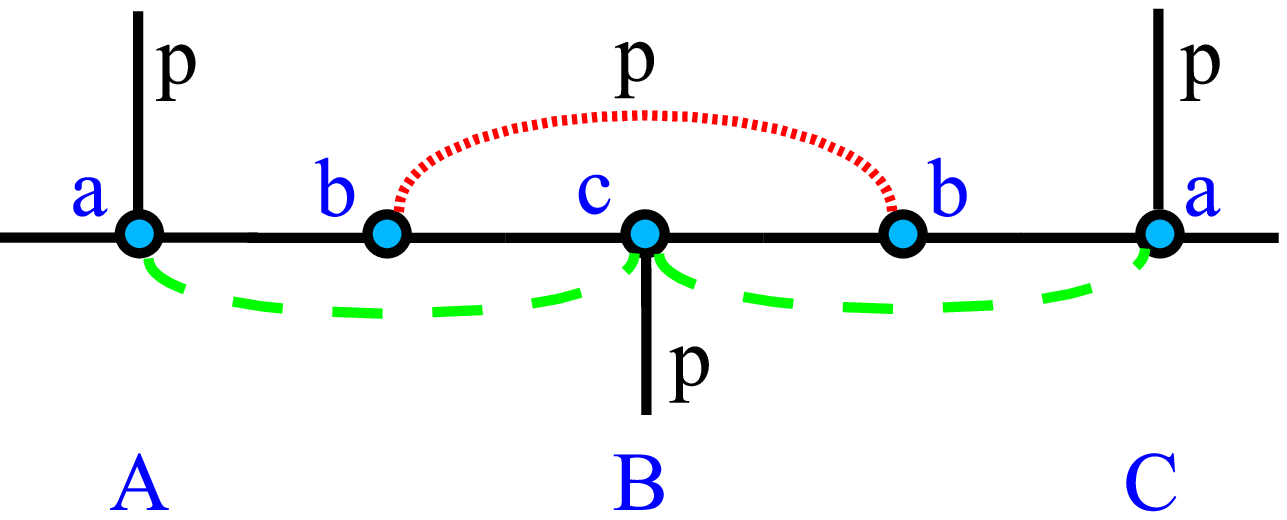}
\includegraphics[bb=70bp 20bp 294bp 200bp,clip,scale=0.5]{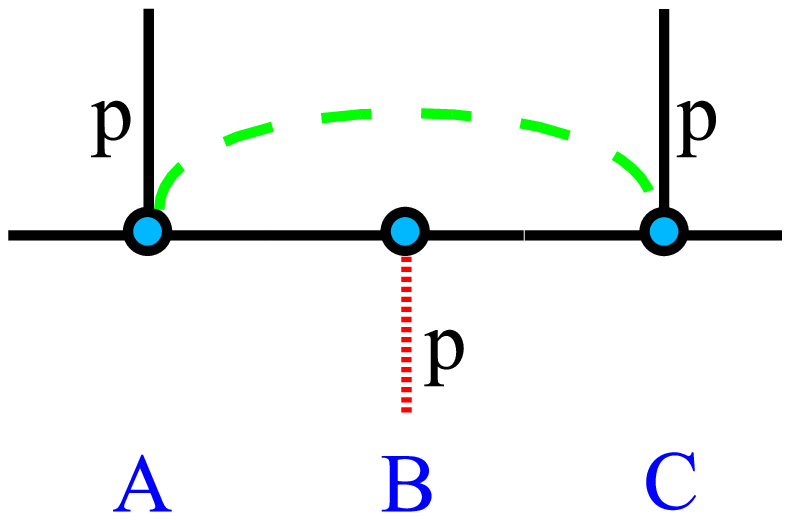}
\caption{\label{fig:RG3}Depiction of the (exact) renormalization group step
for percolation on HN3. The step consists of eliminating every second
site (with label $b$) in the upper plot and expressing the renormalized
connection probabilities $(R_{n+1},S_{n+1},T_{n+1},U_{n+1},N_{n+1})$
on the lower plot in terms of the old values $(R_{n},S_{n},T_{n},U_{n},N_{n})$.
Here, $R_{n}$ refers to the probability that three consecutive points
$abc$ are connected, $S_{n}$ refers to $ab$ being connected but
not $c$, $T_{n}$ to $bc$ being connected but not $a$, $U_{n}$
to $ac$ being connected but not $b$, whereas $N_{n}$ refers to
the probability that neither $a$, $b$, or $c$ are mutually connected.
Upper case letters $A,B,C$ refer to the renormalized sites at $n+1$.
Note that the original network in Fig.~\ref{fig:3hanoi} does not
contain bonds of type $U_{n}$ (long-dashed line), but that they become
relevant during the RG process.}
\end{figure}

First, we consider the renormalization group for end-to-end percolation
on HN3. In this case, all steps can be done exactly. Following Taitelbaum
et al.\ \citep{Taitelbaum90} in their analysis of the Sierpinski gasket,
the RG step consists of eliminating every second (odd) site in the
network and calculating new, renormalized probabilities for connections
between the remaining sites, as explained in Fig.~\ref{fig:RG3}. Four
independent probabilities $(R_{n},S_{n},T_{n},U_{n})$ are evolved under RG;
the  fifth probability, $N_{n}$, follows from conservation of probability.

We obtain a system of RG recursion equations, in which the renormalized
probabilities $(R_{n+1},S_{n+1},T_{n+1},U_{n+1})$ are functions of
$(R_{n},S_{n},T_{n},U_{n};p)$. As for the case of the hierarchical
lattices in Sec.~\ref{sec:MKperco}, these equations are unusual,
as they retain a \emph{memory} of the original bond probability $p$
for all times, which enters as a source-term through the dependence
on the next level in the hierarchy \citep{SWPRL}. The initial conditions
can be read off the right graph-let in Fig.~\ref{fig:RG3}. Unlike
in the other sections, where we merely execute a local stability analysis
near fixed points, here we need to make contact with the initial conditions
in the analysis to explore finite-size corrections. Therefore, we
follow again the notation in Ref.~\citep{Taitelbaum90} and shift
the RG recursion to evolve from initially $n=-k$ to terminate at
$n=0$ for a system of size $L=2^{k}$ in the thermodynamic limit
$k\to\infty$. Originally there is no bond $ac$, thus $U_{-k}=0$
and $R_{-k}=p^{2}$ because it only consists of having both, $ab$
and $bc$, connected. We have $S_{-k}=T_{-k}=p(1-p)$ for an exclusive
$ab$ or $bc$ connection, respectively, hence $N_{-k}=(1-p)^{2}$. 

\begin{figure}[t!]
\includegraphics[bb=0bp 50bp 580bp 720bp,clip,scale=0.45]{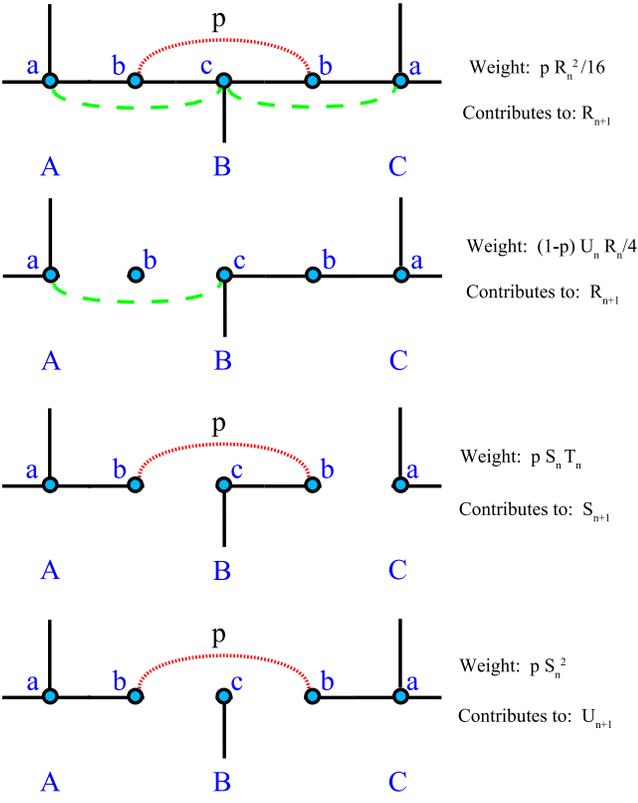}
\caption{\label{fig:RG3counting}Example of graph-lets contributing to the
RG step in HN3. Eliminating every second site in the network (dots
labeled $b$), the connectedness of the renormalized sites $(A,B,C)$
have to be assessed. There are 7 bonds in the elementary graph-let:
2 each for pair-connections $ab$, $bc$, and $ac$, and one previously
unrenormalized bond with raw probability $p$ stretching between the
two sites $b$. Bonds of higher level in the hierarchy (vertical lines)
remain \emph{unaffected} until a later RG step. Thus, $2^{7}=128$
graph-lets with combinations of these bond must be considered. Only
four examples are displayed here. In the first, all bonds are present
with a direct bond between $abc$ on both sides of $B$; each such
triangle has a probability $R_{n}/4$, and the bond $p$ is present,
making the weight $pR_{n}^{2}/16$. Because, after renormalization, the
remaining sites $ABC$ are also directly linked, this graph-let contributes
to $R_{n+1}$. In the second example, only $ac$ but not $b$ are
linked on the left, contributing a weight of $U_{n}$, but there is
a path connecting $abc$ on the right with weight $R_{n}/4$, and
there is no bond $p$, making the total weight $(1-p)U_{n}R_{n}$.
Again, there is a path between $ABC$, so it contributes to $R_{n+1}$.
In the third example, the $S_{n}$ bond between $ab$ on the left
and the $T_{n}$ bond on the right together with the $p$-bond ensure
a connection between $AB$, thus making a contribution to $S_{n+1}$.
The $4th$ example is the only graph-let that contributes to $U_{n+1}$
with a direct link between $AC$ but not $B$. }
\end{figure}

We determine the form of the recursion equations for these probabilities
by a simple counting procedure, as explained in Fig.~\ref{fig:RG3counting}.
We find
\begin{eqnarray}
R_{n+1} & = & R_{n}^{2}+2U_{n}R_{n}+U_{n}^{2}+2pS_{n}R_{n},\nonumber \\
S_{n+1} & = & R_{n}T_{n}+R_{n}N_{n}+U_{n}N_{n}+U_{n}T_{n}\nonumber \\
 &  & \quad+U_{n}S_{n}+pS_{n}T_{n}+(1-p)R_{n}S_{n},\nonumber\\
T_{n+1} & = & R_{n}T_{n}+R_{n}N_{n}+U_{n}N_{n}+U_{n}T_{n}\nonumber \\
 &  & \quad+U_{n}S_{n}+pS_{n}T_{n}+(1-p)R_{n}S_{n},\nonumber\\
U_{n+1} & = & pS_{n}^{2}, \label{eq:RGrecur} \\
N_{n+1} & = & T_{n}^{2}+2N_{n}S_{n}+2N_{n}T_{n}+N_{n}^{2}\nonumber\\
 &  & \quad+2(1-p)S_{n}T_{n}+(1-p)S_{n}^{2}.\nonumber
\end{eqnarray}
Note that unlike for the Sierpinski gasket \citep{Taitelbaum90} but
similar to Eqs.~(\ref{eq:RGrecurMK1}) and~(\ref{eq:RGrecurMK2}),
the RG recursions depend explicitly on $p$ at every step, not merely
through the initial conditions. 

Obviously, the equations imply that $T_{n}=S_{n}$ for all $n$, making
the $T$-operator redundant. Combining probabilities $S_{n}$ and
$T_{n}$ into a single probability $S_{n}$ for having a connection
between either $ab$ or $bc$, and rescaling by a factor of $\frac{1}{2}$,
Eqs.~(\ref{eq:RGrecur}) can be simplified to
\begin{eqnarray}
R_{n+1} & = & \left(R_{n}+U_{n}\right)^{2}+pS_{n}R_{n},\nonumber \\
S_{n+1} & = & 2\left(S_{n}+U_{n}\right)\left(R_{n}+U_{n}\right)+\frac{p}{2}S_{n}^{2}-pR_{n}S_{n},\nonumber \\
U_{n+1} & = & \frac{1}{4}pS_{n}^{2},\label{eq:NewRGrecur}\\
N_{n+1} & = &
\left(S_{n}+N_{n}\right)^{2}-\frac{3}{4}pS_{n}^{2},\nonumber 
\end{eqnarray}
which satisfies the constraint
\begin{eqnarray}
1 & = & N_{n}+R_{n}+S_{n}+U_{n}\label{eq:Norm}
\end{eqnarray}
 $ $for all $n$. The initial conditions at $n=-k$ are now
\begin{eqnarray}
R_{-k} & = & p^{2},\nonumber \\
S_{-k} & = & 2p(1-p),\nonumber \\
U_{-k} & = & 0,\label{eq:RGinit}\\
N_{-k} & = & (1-p)^{2}.\nonumber 
\end{eqnarray}

Note that the fixed point equations can be solved exactly for arbitrary
$p$, as they involve the solution of three coupled quadratic equations, but
the expressions are extremely complex. Instead, we search for fixed
points of these equations numerically. For any physical $p$ inside
the unit interval we find  \emph{no} physical fixed points with $0<N^{*},R^{*},S^{*},U^{*}<1$.

The analysis of the RG recursion Eqs.~(\ref{eq:NewRGrecur}) proceeds
as follows. The obligatory trivial fixed points, i.~e., stationary
solutions of Eqs.~(\ref{eq:NewRGrecur}), at $\left(N^{*},R^{*},S^{*},U^{*}\right)=(1,0,0,0)$
and $(0,1,0,0)$ exist for any $p$. We search for a non-trivial fixed
point by choosing a value of $0<p<1$, and numerically evolve Eqs.
(\ref{eq:NewRGrecur}) for that value of $p$, starting from the initial
conditions in (\ref{eq:RGinit}). We find that for all values of $p$,
the equations evolve towards the stable fixed point at $(1,0,0,0)$,
characteristic for a completely disconnected network, although this evolution
takes ever longer, the closer $p$ is to unity. Therefore, we
conclude:
\begin{eqnarray}
p_{c} & = & 1.
\label{eq:pc}
\end{eqnarray}
This result is not too surprising, because this network is of finite
ramification: a bounded number of cuts can separate the network into
two extensive pieces. Finitely ramified objects, such as the triangular
Sierpinski gasket, can only percolate with certainty at full connectivity,
$p_{c}=1$. Corrections to this behavior in HN3 are more interesting
and, in fact, decay \emph{faster} for $k\to\infty$ than for other
finitely ramified objects like the Sierpinski gasket.

\subsubsection{First-Order Correction to $p_{c}=1$:\label{sub:First-Order-Correction-to}}
To determine the finite-size corrections for $p_{c}=1$, we expand
around the \emph{unstable} fixed point $(0,1,0,0)$ to first order
with the Ansatz
\begin{eqnarray}
R_{n} & \sim & 1-r_{n}^{(1)},\quad\left(r_{n}^{(1)}\ll1\right),\nonumber \\
S_{n} & \sim & s_{n}^{(1)},\qquad\left(s_{n}^{(1)}\ll1\right),\nonumber \\
U_{n} & \sim & u_{n}^{(1)},\qquad\left(u_{n}^{(1)}\ll1\right),\label{eq:Ansatz}\\
N_{n} & \sim &
\eta_{n}^{(1)},\qquad\left(\eta_{n}^{(1)}\ll1\right).\nonumber 
\end{eqnarray}
Inserting into Eqs.~(\ref{eq:NewRGrecur}) and dropping terms beyond
first order, we obtain immediately that
\begin{eqnarray}
u_{n}^{(1)} & = & \eta_{n}^{(1)}=0.
\end{eqnarray}
The norm in Eq.~(\ref{eq:Norm})  implies
$1\sim1-r_{n}^{(1)}+s_{n}^{(1)}$, i.~e.,
\begin{eqnarray}
r_{n}^{(1)} & = & s_{n}^{(1)}
\label{eq:r1eqs1}
\end{eqnarray}
and we get from the equations in (\ref{eq:NewRGrecur}) for $S_{n+1}$:
\begin{eqnarray}
s_{n+1}^{(1)} & \sim & \left(2-p\right)s_{n}^{(1)},
\label{eq:rs}
\end{eqnarray}
which together with Eq.~(\ref{eq:r1eqs1}) have the consistent solution
\begin{eqnarray}
r_{n}^{(1)} & = & s_{n}^{(1)}\sim A(2-p)^{n}.
\label{eq:rn}
\end{eqnarray}
Applying the initial conditions in (\ref{eq:RGinit}), we can now
assess the finite-size corrections to $p_{c}=1$ with the Ansatz
\begin{eqnarray}
p & \sim & 1-\epsilon_{k},\qquad\left(\epsilon_{k}\ll1\right).
\label{eq:pAnsatz}
\end{eqnarray}
Inserting this and the result for $r_{-k}^{(1)}$ and $s_{-k}^{(1)}$
in Eq.~(\ref{eq:rn}) into Eqs.~(\ref{eq:RGinit}) yields a consistent
result, which is
\begin{eqnarray}
A\left(1+\epsilon_{k}\right)^{k} & \sim & 2\epsilon_{k}.
\label{eq:epsilonk}
\end{eqnarray}
Note also that with $p$ in Eq.~(\ref{eq:pAnsatz}), the last initial
condition in Eq.~(\ref{eq:RGinit}) evaluates consistently to $N_{-k}=O\left(\epsilon_{k}^{2}\right)$. 

To leading order, we obtain from Eq.~(\ref{eq:epsilonk}) 
\begin{eqnarray}
\epsilon_{k} & \sim & \frac{\ln k}{k},
\label{eq:firstcorrection}
\end{eqnarray}
independent of the unknown constant $A$. In general it would be best
to extract $\epsilon_{k}$ directly from Eq.~(\ref{eq:epsilonk})
for any given $k$, as Eq.~(\ref{eq:firstcorrection}) is poorly
convergent.
Unfortunately, though, we have not found a way to fix $A$ in this
local expansion. We conclude here by noting that the leading corrections
to $p_{c}=1$ as stated in Eq.~(\ref{eq:firstcorrection}) decays
faster than what Ref.~\citep{Taitelbaum90} found for the Sierpinski
gasket, $\epsilon_{k}\sim1/\left(2\sqrt{k}\right)$, where $k=\log_{2}N$.
We continue with the second-order correction in the Appendix, with the
final result given in Eq.~(\ref{eq:pc2order}).

\begin{figure}
\includegraphics[bb=0bp 20bp 369bp 200bp,clip,scale=0.5]{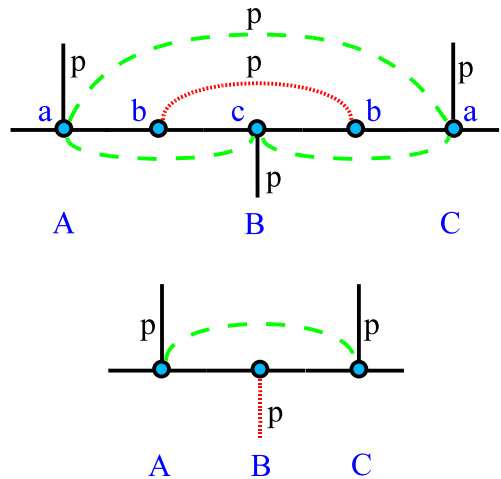}
\includegraphics[bb=70bp 20bp 294bp 200bp,clip,scale=0.5]{RG3perco_after}
\caption{\label{fig:RG5}Depiction of the (exact) RG step
for percolation on HN5. In contrast to the corresponding step for
HN3 in Fig.~\ref{fig:RG3}, the long-dashed bonds are germane to the
network, not emergent properties. Thus, the long-dashed bond on top
has still probability $p$; the long-dashed bonds of the previous
hierarchical level below have a probability of $p$ plus the emergent
contribution from the previous step. The graph-let after
the RG step (bottom) is identical to that in Fig.~\ref{fig:RG3}.}
\end{figure}

\subsection{RG for HN5\label{sub:RG_HN5}}
Just as HN5 is a simple extension of HN3, see Sec.~\ref{sec:Graph-Structure}
above, so is the analysis of the percolation problem. Yet, the added
bonds in HN5 precisely rectify the finite ramification that made the
percolation phase diagram for HN3 trivial. On the face of it, it is
not entirely obvious that percolation on HN5 would be any more interesting,
because its ramification is only marginally infinite. Now the number
of lines to be cut grows logarithmically with size $ $$N=2^{k}$,
because each new level $k$ of the hierarchy adds another end-to-end
connection, see Fig.~\ref{fig:5hanoi}.

The RG step for HN5 involves merely one extra bond in the elementary
graph-let, see Fig.~\ref{fig:RG5}.  The derivation of
the RG recursion equations is similar to Eq.~(\ref{eq:NewRGrecur}), however,
we now have to evaluate twice as many graph-lets. But one half of the graph-lets---those
without the new line---simply correspond exactly to those of HN3 in
weight (times a factor of $1-p$) and in the operator to which they
contribute. The other half again have the same weight (times a factor
$p$, for the now-present new bond), but the additional bond changes
the operator to which these graph-lets contribute in 69 out of the
128 cases by adding extra connectivity. These considerations and the
corresponding discussion along the lines of Sec.~\ref{sub:RG_HN3}
lead to
\begin{eqnarray}
R_{n+1} & = & \left(R_{n}+U_{n}\right)^{2}+2p\left(S_{n}+N_{n}\right)\left(R_{n}+U_{n}\right)\nonumber \\
 &  & \quad+p\left(1-p\right)S_{n}R_{n}+\frac{p^{2}}{2}S_{n}^{2},\nonumber\\
S_{n+1} & = & 2\left(1-p\right)\left[\left(S_{n}+N_{n}\right)\left(R_{n}+U_{n}\right)+\frac{p}{4}S_{n}^{2}\right]\nonumber \\
 &  & \quad-p\left(1-p\right)R_{n}S_{n},\nonumber\\
U_{n+1} & = & p\left[\left(S_{n}+N_{n}\right)^{2}+\frac{1-3p}{4}S_{n}^{2}\right],\label{eq:HN5RGrecur}\\
N_{n+1} & = &
\left(1-p\right)\left[\left(S_{n}+N_{n}\right)^{2}-\frac{3}{4}pS_{n}^{2}\right],\nonumber 
\end{eqnarray}
which again satisfies the norm-constraint in Eq.~(\ref{eq:Norm}).
The initial conditions are now\begin{eqnarray}
R_{0} & = & p^{2}\left(3-2p\right),\nonumber \\
S_{0} & = & 2p(1-p)^{2},\nonumber \\
U_{0} & = & p\left(1-p\right)^{2},\label{eq:HN5RGinit}\\
N_{0} & = & (1-p)^{3}.\nonumber 
\end{eqnarray}

\begin{figure}
\includegraphics[clip,scale=0.3]{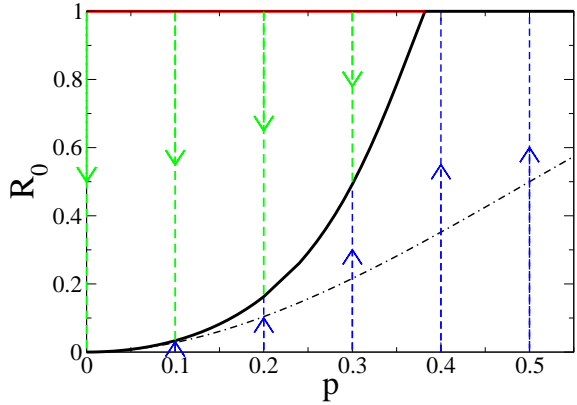}
\caption{\label{fig:HN5phase}Phase diagram for the RG flow of HN5 due to Eq.~(\ref{eq:HN5RGrecur}).
The all-connected operator $R$ is used as an indicator of percolative
order in the system. For any given $p$, $R_{n}$ evolves along the
shown trajectories starting from an initial value $R_{0}$. Except
for points $\left(p<\bar{p}=2-\phi,R_{0}=1\right)$ on the (thick
red) line of unstable fixed points, all points $\left(p,R_{0}<1\right)$
flow towards the (thick black) line of stable fixed points obtained
from Eq.~(\ref{eq:HN5RGrecur}) or to $R^{*}=1$ for $p>\bar{p}$.
For a homogeneous choice of bond probabilities in the network given
in Eqs.~(\ref{eq:HN5RGinit}), i.~e., $R_{0}=p^{2}(3-2p)$ along the
dash-dotted line, the stable fixed point is always approached
from below.}
\end{figure}

The algebraic fixed-point equations obtained from
Eqs.~(\ref{eq:HN5RGrecur}) for $n\approx n+1\to\infty$ are extremely
complex. It is easy to establish the fixed point at
$\left(N^{*},R^{*},S^{*},U^{*}\right)=(0,1,0,0)$ for any $p$. But the
other trivial stable fixed point at
$\left(N^{*},R^{*},S^{*},U^{*}\right)=(1,0,0,0)$, describing the
disordered state, \emph{only} exists at $p=0$. Instead, we find
(numerically, as the solution of a sixth-order polynomial) a
nontrivial equation for a line of fixed-point which merges into the
fully ordered, percolating fixed-point $R^{*}=1$ that is stable for
all $p>\bar{p}=2-\phi=0.381966\ldots$, where
$\phi=\left(\sqrt{5}+1\right)/2$.  The resulting phase diagram in
Fig.~\ref{fig:HN5phase} is very similar to that in
Fig.~\ref{fig:MK1phase} for the hierarchical small-world lattice. The
difference is that HN5 is more mean-field like in the sense of
ordinary random graphs, with an exponential instead of a scale-free
degree distribution.

We obtain the value for $\bar{p}$ explicitly from the analysis of
the fixed point at $R^{*}=1$. In fact, we proceed similar to Sec.
\ref{sub:RG_HN3} by applying the same Ansatz in Eq.~(\ref{eq:AnsatzR0})
to the RG recursions in (\ref{eq:HN-NPRGrecur}). A first-order expansion yields
\begin{eqnarray}
r_{n}^{(1)} & = & s_{n}^{(1)},\nonumber \\
s_{n+1}^{(1)} & \sim &
\left(2-p\right)\left(1-p\right)s_{n}^{(1)},\label{eq:rsHN5}
\end{eqnarray}
with $u_{n}^{(1)}=\eta_{n}^{(1)}\equiv0$. Finally, we have
\begin{eqnarray}
r_{n}^{(1)} & = & s_{n}^{(1)}\sim
A\left[\left(2-p\right)\left(1-p\right)\right]^{n}.
\label{eq:rs_solutionHN5}
\end{eqnarray}
Despite the obvious similarity to Eqs.~(\ref{eq:AnsatzR0}-\ref{eq:rn}),
the key difference of the solution in Eq.~(\ref{eq:rs_solutionHN5})
is that there is a non-trivial transition at $\bar{p}=2-\phi$, derived
from the marginal stability condition, 
\begin{eqnarray}
1 & = & \left(2-\bar{p}\right)\left(1-\bar{p}\right),
\label{eq:pbarHN5}
\end{eqnarray}
such that the fixed point is unstable for $0\leq p<\bar{p}$ and stable
for $\bar{p}<p\leq1$. Clearly, the approach to the fixed point is
highly parameter-dependent, leading to non-universality such as in
the correlation-length exponent for $p<\bar{p}$: \citep{Plischke94}
\begin{eqnarray}
\nu & = &
\frac{\ln2}{\ln\left[\left(2-p\right)\left(1-p\right)\right]}.
\label{eq:nuHN5}
\end{eqnarray}

\begin{figure}
\includegraphics[bb=0bp 17bp 369bp 200bp,clip,scale=0.5]{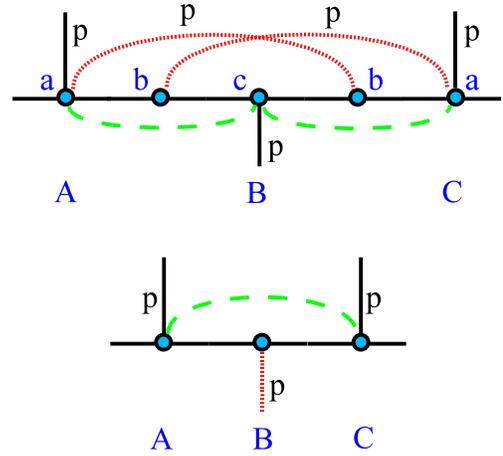}
\includegraphics[bb=70bp 20bp 294bp 200bp,clip,scale=0.5]{RG3perco_after}
\caption{\label{fig:RGNP}Depiction of the (exact) RG step
for percolation on HN-NP. In contrast to HN3, the long-range bond
connecting $bb$ in Fig.~\ref{fig:RG3} is replace by a crossing pair
of such bonds connecting $ab$. The $U$-bond connecting $ac$ is an emergent operator,
like in HN3.  The graph-let after the
RG step (bottom) is identical to that for HN3 in Fig.~\ref{fig:RG3}
and HN5 in Fig.~\ref{fig:RG5}.}
\end{figure}

\subsection{Percolation in HN-NP\label{sub:RG_HN-NP}}
A more substantial extension to the previous planar Hanoi networks
is given by HN-NP introduced in Sec.~\ref{sec:Introduction}, see
Fig.~\ref{fig:hanoi_nonplanar}. It extends the basic network-building
concept into the non-planar regime while preserving exact renormalizability.
The single, long-range bond of HN3 in the elementary graph-let in
Fig.~\ref{fig:RG3} is here replaced by a pair of crossing small-world
bonds, as explained in Fig.~\ref{fig:RGNP}. As the second diagram
shows, the graph-let resulting from the RG step is \emph{identical}
to those for HN3 and HN5. As in HN3, only the $U$-operator that connect
sites $ac$ exclusively emerges anew during the RG recursion. (Elsewhere,
we also consider that operator as pre-existing for HN-NP, the same
step that converts HN3 into HN5 \citep{HanoiIsing09}.) 

Evaluation of the $2^{8}=256$ elementary graph-lets as described
in connection with Fig.~\ref{fig:RG3counting} above results in the
recursion equations
\begin{eqnarray}
R_{n+1} & = & \left(R_{n}+U_{n}\right)^{2}+p\left(3-p\right)R_{n}S_{n}+2pR_{n}N_{n}\nonumber \\
 &  & \quad+pS_{n}U_{n}+\frac{3}{4}p^{2}S_{n}^{2},\nonumber\\
S_{n+1} & = & \left(2-p\right)\left[\left(1-p\right)R_{n}+U_{n}\right]S_{n}+2\left(1-p\right)R_{n}N_{n}\nonumber \\
 &  & \quad+2U_{n}N_{n}+pN_{n}S_{n}+p\left(1-p\right)S_{n}^{2},\nonumber\\
U_{n+1} & = & p\left(1-\frac{3}{4}p\right)S_{n}^{2}+pN_{n}S_{n},\label{eq:HN-NPRGrecur}\\
N_{n+1} & = &
N_{n}^{2}+2\left(1-p\right)S_{n}N_{n}+\left(1-p\right)^{2}S_{n}^{2},\nonumber 
\end{eqnarray}
obeying the normalization constraint in Eq.~(\ref{eq:Norm}), and
with the same initial conditions as for HN3 in Eqs.~(\ref{eq:RGinit}).

\begin{figure}
\includegraphics[clip,scale=0.3]{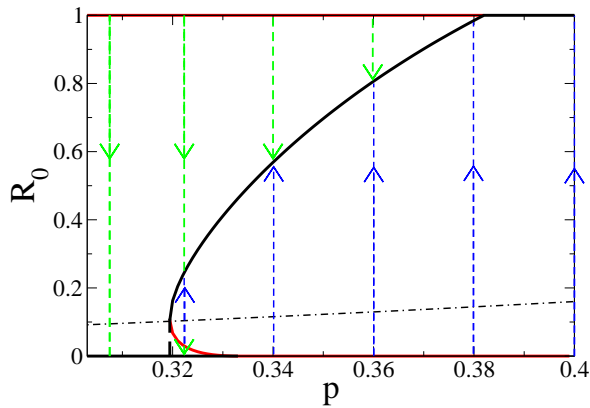}
\caption{\label{fig:HN-NPphase}Phase diagram for the RG flow of HN-NP due to Eq.~(\ref{eq:HN-NPRGrecur}).
Here, the all-connected operator $R$ is used as in indicator of percolative
order in the system. For any given $p$, $R_{n}$ evolves along the
shown trajectories starting from an initial value $R_{0}$. The homogeneous
choice of bond probabilities in the network given in Eqs.~(\ref{eq:RGinit}),
i.~e., $R_{0}=p^{2}$ along the (dash-dotted) diagonal line, crosses
the line of unstable fixed points \emph{just below} the branch-point
singularity.}
\end{figure}

At first glance, the phase diagram for HN-NP appears to have {}``cured''
all the unusual small-world effects of the previous networks and reverted
to ordinary mean-field-like behavior: There is a domain for small
$p\geq0$ where the percolation probability vanishes and one at larger
$p\leq1$ where percolation is certain, with what seems to be the
usual 0-1 transition in the probability at some critical $p_{c}$
(see, e.~g., Fig.~17 in Ref.~\citep{Stauffer94}). Yet, closer inspection
of a relatively narrow regime near that transition, about for $0.3<p<0.4$,
reveals a more subtle behavior. If we didn't have the exact RG equations
in (\ref{eq:HN-NPRGrecur}) available, a numerical simulation may
well have missed it due to finite-size corrections. Within that narrow
region, enlarged in Fig.~\ref{fig:HN-NPphase}, the resulting phase
diagram is similar to Fig.~\ref{fig:MK2phase}, with a
branch-point  where two lines of stable and unstable
fixed points merge and disappear into the complex domain. But this
branch-point is oriented in the opposite direction, with a (minute)
jump in the percolation probability at $\bar{p_{l}}=0.319445181\ldots$
preceding a continuous rise towards, again, $\bar{p}_{u}=2-\phi=0.381966\ldots$
above which percolation is certain. Unlike for Fig.~\ref{fig:MK2phase},
$R^{*}=0$ and $R^{*}=1$ are \emph{always} fixed points (variably
stable or unstable) for all $p$: While in the $2d$ hierarchical
lattice the small-world bond for any $p$ can boot-strap even the
empty network into a spanning cluster with finite probability, in
HN-NP this is only possible for a disconnected network when $p>\bar{p}_{m}=\frac{1}{3}$.

The analysis near the fixed points provides more details about these
observations. The results for the fixed point at $R^{*}=1$ for $p<\bar{p}_{u}$
is identical to that for HN5 in Sec.~\ref{sub:RG_HN5}, see Eq.~(\ref{eq:rsHN5}),
and provides again $\bar{p}_{u}=2-\phi$ as the transition
point between the line of unstable and stable fixed points. In turn,
the analysis near $R^{*}=0$, is novel. Enforcing the norm-constraint
$N_{n}=1-R_{n}-S_{n}-U_{n}$ and expanding around the unstable fixed
point to first order with the Ansatz
\begin{eqnarray}
R_{n} & \sim & r_{n}^{(1)},\quad\left(r_{n}^{(1)}\ll1\right),\nonumber \\
S_{n} & \sim & s_{n}^{(1)},\qquad\left(s_{n}^{(1)}\ll1\right),\nonumber \\
U_{n} & \sim &
u_{n}^{(1)},\qquad\left(u_{n}^{(1)}\ll1\right),\label{eq:AnsatzR0}
\end{eqnarray}
leads to 
\begin{eqnarray}
r_{n+1}^{(1)} & \sim & 2pr_{n}^{(1)},\nonumber \\
s_{n+1}^{(1)} & \sim & 2\left(1-p\right)r_{n}^{(1)}+ps_{n}^{(1)}+2u_{n}^{(1)},\nonumber \\
u_{n+1}^{(1)} & \sim & ps_{n}^{(1)}.\label{eq:HN-NPorder1}
\end{eqnarray}
The first relation immediately yields
\begin{eqnarray}
r_{n}^{(1)} & \sim & \left(2p\right)^{n}r_{0}^{(1)},
\label{eq:HN-NPr1}
\end{eqnarray}
and the physically relevant solution for the remaining coupled set
of linear recursions provides
\begin{eqnarray}
s_{n}^{(1)} & \sim & u_{n}^{(1)}\sim
A\left[\frac{p}{2}\left(1+\sqrt{1+\frac{8}{p}}\right)\right]^{n}.
\label{eq:HN-NPeigenvalue}
\end{eqnarray}
(The modulus of the second, oscillatory, solution decays with
$n$ for all $0\leq p\leq1$.) Again, the highly non-trivial scaling
exponent derived from Eq.~(\ref{eq:HN-NPeigenvalue}) is non-universal.
For small $p<\bar{p}_{m}=\frac{1}{3}$, all corrections decay and
the fixed point remains stable. For $p>\bar{p}_{m}$, the solutions
for $s_{n}^{(1)}$ and $u_{n}^{(1)}$ first become divergent, enough
to make the fixed point itself unstable. Yet, $r_{n}^{(1)}$ initially
remains convergent, slowing the RG flow. For $p>\frac{1}{2}$, the
fixed point rapidly becomes unstable. The analysis of the branch-point
at $\bar{p}_{l}$ can only be done numerically because it involves
the solution of sufficiently high-order polynomials to defy an explicit
treatment.

\section{Conclusions\label{sec:Conclusions}}
We have obtained the exact phase diagrams for bond-percolation on
variants of the Hanoi networks using RG. For a small fraction of long-range
bonds and a regular degree distribution in HN3, which leaves the network
finitely ramified and with an average distance that grows like a square-root
of the system size, percolation requires a full bond-density, $p_{c}=1$.
The hierarchical structure there manifests itself only in the finite-size
corrections. Adding more long-range bonds at the price of obtaining
an exponential degree distribution but a mean-field like logarithmic
distance between sites results in the network HN5. There, it leads
to the unusual phenomenon of percolation with a finite probability,
which we dubbed {}``patchy'' order, facilitated by the addition
of ever more long-range bonds at higher levels of the hierarchy that
can reconnect patches of localized clusters at lower levels of the
hierarchy. Introducing a non-planar version of a Hanoi graph provides
nearly ideal mean-field behavior, with a completely disordered phase
at low bond-density and a fully ordered phase at high density. Yet,
even in this case a narrow window with mixed behavior intervenes. 

Here, we have all but scratched the surface of the analysis of percolation
on small-world networks. In light of the ever increasing importance
of small-world networks, practical and conceptual, we will focus on
exploring further characteristics, such as details of clustering behaviors
and correlations, on the variety of network configurations represented
here. Only then can we hope to find solid classifications on the possible
behaviors found in the real world. For instance, it might be possible
to find recently observed discontinuous transitions in the size of
spanning clusters ({}``explosive'' percolation \citep{Achlioptas09,Ziff09,Cho09,Radicchi09}
also in the neighborhood of the branch points of these merely hierarchically
constructed networks.

\section*{Appendix\label{sec:Appendix}}
\subsection*{Second-Order Correction to $p_{c}=1$ in HN3:\label{sub:Second-Order-Correction-to}}
We expand around the unstable fixed point $(0,1,0,0)$ of the RG recursions
(\ref{eq:NewRGrecur}) for HN3 to second order with the Ansatz
\begin{eqnarray}
R_{n} & \sim & 1-r_{n}^{(1)}+r_{n}^{(2)},\quad\left(r_{n}^{(2)}\ll r_{n}^{(1)}\ll1\right),\nonumber \\
S_{n} & \sim & r_{n}^{(1)}+s_{n}^{(2)},\qquad\left(s_{n}^{(2)}\ll r_{n}^{(1)}\ll1\right),\nonumber \\
U_{n} & \sim & u_{n}^{(2)},\qquad\left(u_{n}^{(2)}\ll r_{n}^{(1)}\ll1\right),\label{eq:Ansatz2}\\
N_{n} & \sim & \eta_{n}^{(2)},\qquad\left(\eta_{n}^{(2)}\ll
r_{n}^{(1)}\ll1\right),\nonumber 
\end{eqnarray}
where we have used Eq.~(\ref{eq:r1eqs1}) to eliminate $s_{n}^{(1)}$
in favor of $r_{n}^{(1)}$. Of course, all terms of lower order cancel,
and after dropping terms of higher than second order, we arrive at
\begin{eqnarray}
r_{n+1}^{(2)} & \sim & 2r_{n}^{(2)}+2u_{n}^{(2)}+ps_{n}^{(2)}+\left(1-p\right)\left(r_{n}^{(1)}\right)^{2},\nonumber \\
s_{n+1}^{(2)} & \sim & \left(2-p\right)s_{n}^{(2)}+2\eta_{n}^{(2)}+\left(\frac{3}{2}p-2\right)\left(r_{n}^{(1)}\right)^{2},\nonumber \\
u_{n+1}^{(2)} & \sim & \frac{p}{4}\left(r_{n}^{(1)}\right)^{2},\label{eq:2order}\\
\eta_{n+1}^{(2)} & \sim &
\left(1-\frac{3}{4}p\right)\left(r_{n}^{(1)}\right)^{2},\nonumber 
\end{eqnarray}
assuming that squares of terms of first order are of second order.
The solutions for $u_{n}^{(2)}$ and $\eta_{n}^{(2)}$ are 
\begin{eqnarray}
u_{n}^{(2)} & = & \frac{p}{4}\, A^{2}\left(2-p\right)^{2n-2},\nonumber \\
\eta_{n}^{(2)} & = & \left(1-\frac{3}{4}p\right)\,
A^{2}\left(2-p\right)^{2n-2}.
\label{eq:u-eta2}
\end{eqnarray}
Inserting these results into the equation for $s_{n+1}^{(2)}$ yields
\begin{eqnarray}
s_{n+1}^{(2)} & \sim & \left(2-p\right)s_{n}^{(2)}\\
 &  &\quad+\left(2-\frac{3}{2}p\right)\left[1-\left(2-p\right)^{2}\right]\,
A^{2}\left(2-p\right)^{2n-2}.\nonumber
\end{eqnarray}
The inhomogeneous term in this relation can be neglected at this order,
in fact, as the expression in the square brackets vanishes in the
limit $p\to1$, and we are left with the same solution for $s_{n}^{(2)}$
as in first order,
\begin{eqnarray}
s_{n}^{(2)} & = & B\left(2-p\right)^{n},
\label{eq:s2}
\end{eqnarray}
where $B\ll A$.$ $ (In fact, we expect $B/A\sim\epsilon_{k}$.)

Regarding the relation for $r_{n}^{(2)}$ in Eq.~(\ref{eq:2order}),
we can drop the last term in this order as $1-p\to0$ and find
\begin{eqnarray}
r_{n+1}^{(2)} & \sim & 2r_{n}^{(2)}+2u_{n}^{(2)}+ps_{n}^{(2)},\nonumber \\
 & \sim & 2r_{n}^{(2)}+\frac{p}{2}\, A^{2}\left(2-p\right)^{2n-2}+p\,
B\left(2-p\right)^{n}.
\label{eq:rn2inhom}
\end{eqnarray}
The homogeneous solution of this relation at $n=-k$ would decay exponentially,
$\sim2^{-k}$, and can be discarded, leaving us with
\begin{eqnarray}
r_{n}^{(2)} & \sim & \frac{p}{2}\, A^{2}\left(2-p\right)^{2n-4}+p\,
B\left(2-p\right)^{n+1}.
\label{eq:rn2solution}
\end{eqnarray}

Inserting the second-order solutions in Eqs.~(\ref{eq:u-eta2},\ref{eq:s2},\ref{eq:rn2solution})
together with the results from Sec.~\ref{sub:First-Order-Correction-to}
into Eqs.~(\ref{eq:Ansatz2}) and evaluating at $g=k$, we get
\begin{eqnarray}
R_{-k} & \sim & 1-A\left(2-p\right)^{-k}+p\, B\left(2-p\right)^{-k-1}\nonumber \\
 &  & \quad+\frac{p}{2}\, A^{2}\left(2-p\right)^{-2k-4},\nonumber\\
S_{-k} & \sim & A\left(2-p\right)^{-k}+B\left(2-p\right)^{-k},\qquad(A\gg B),\nonumber \\
U_{-k-1} & \sim & \frac{p}{4}\, A^{2}\left(2-p\right)^{-2k},\label{eq:at-k}\\
N_{-k} & \sim &
\left(1-\frac{3}{4}p\right)A^{2}\left(2-p\right)^{-2k-2}.\nonumber 
\end{eqnarray}
We extend the correction on $p_{c}$ to second order, i.~e.,
\begin{eqnarray}
p & \sim & 1-\epsilon_{k}+b\,\epsilon_{k}^{2},
\label{eq:p2order}
\end{eqnarray}
where $b=b(\epsilon_{k})$ should only vary weakly with $\epsilon_{k}$.
Inserted into Eqs.~(\ref{eq:at-k}), we get
\begin{eqnarray}
R_{-k} & \sim & 1-2\epsilon_{k}+\epsilon_{k}^{2}\left[2+2b\ln\left(\frac{2\epsilon_{k}}{A}\right)+2\frac{B}{\epsilon_{k}A}\right],\nonumber \\
S_{-k} & \sim & 2\epsilon_{k}+\epsilon_{k}^{2}\left[-2b\ln\left(\frac{2\epsilon_{k}}{A}\right)+2\frac{B}{\epsilon_{k}A}\right],\nonumber \\
U_{-k+1} & \sim & \epsilon_{k}^{2},\label{eq:all-kinepsilon}\\
N_{-k} & \sim & \epsilon_{k}^{2},\nonumber 
\end{eqnarray}
where we used 
\begin{eqnarray}
A\left(2-p\right)^{-k} & \sim & A\left(1+\epsilon_{k}-b\epsilon_{k}^{2}\right)^{-k},\nonumber\\
 & \sim & A\left(1+\epsilon_{k}\right)^{-k}\left[1+bk\epsilon_{k}^{2}\right],\\
 & \sim &
2\epsilon_{k}\left[1-b\ln\left(\frac{2\epsilon_{k}}{A}\right)\epsilon_{k}\right],\nonumber
\end{eqnarray}
because Eq.~(\ref{eq:epsilonk}) yields $-k\epsilon_{k}\sim\ln\left(2\epsilon_{k}/A\right)$. 

Correspondingly, we can expand the initial conditions 
\begin{eqnarray}
R_{-k} & = & p^{2}\nonumber \\
 & \sim & 1-2\epsilon_{k}+\epsilon_{k}^{2}\left[1+2b\right],\nonumber \\
S_{-k} & = & 2p(1-p)\label{eq:epsilon_init}\\
 & \sim & 2\epsilon_{k}+\epsilon_{k}^{2}\left[-2-2b\right],\nonumber \\
N_{-k} & = & (1-p)^{2}\nonumber \\
 & \sim & \epsilon_{k}^{2}.\nonumber 
\end{eqnarray}
Note that in Eq.~(\ref{eq:all-kinepsilon}) we have used the result
for the newly emerging operator $U$ at $g=k-1$, because $U_{-k}$
itself would strictly vanish to all orders. But with the help of its
relation in Eq.~(\ref{eq:NewRGrecur}), we obtain via the initial
condition for $S_{n}$:
\begin{eqnarray}
U_{-k+1} & = & p^{3}\left(1-p\right)^{2}\label{eq:Uinit}\\
 & \sim & \epsilon_{k}^{2}.\nonumber 
\end{eqnarray}

Comparison of those results with Eqs.~(\ref{eq:all-kinepsilon}) prove
the consistency for $ $$N$ and $U$. As expected from Sec.~\ref{sub:First-Order-Correction-to},
the relations for $R$ and $S$ are also consistent immediately up
to first order. Requiring consistency to second order provides
relations
\begin{eqnarray}
1+2b & = & 2+2b\ln\left(\frac{2\epsilon_{k}}{A}\right)+2\frac{B}{\epsilon_{k}A},\nonumber\\
-2-2b & = &
-2b\ln\left(\frac{2\epsilon_{k}}{A}\right)+2\frac{B}{\epsilon_{k}A},\nonumber
\end{eqnarray}
that determine the integration constant $B$ and, more importantly,
the second-order correction $b$ to $p_{c}$:
\begin{eqnarray}
b & = & -\frac{1}{4}\,\frac{1}{1+\ln\left(\frac{A}{2\epsilon_{k}}\right)},\label{eq:bsolution}\\
B & = & -\frac{3}{4}\epsilon_{k}A.\nonumber 
\end{eqnarray}
A redundant test of these results is provided by the normalization
constraint, 
\begin{eqnarray}
1 & = & R_{-k}+S_{-k}+N_{-k},
\end{eqnarray}
which we find satisfied to second order. (Remember that $U_{-k}=0$.) 

Hence, we obtain finally for the finite-size corrections of $p_{c}$:
\begin{eqnarray}
p_{c} & \sim & 1-\epsilon_{k}-\frac{1}{4}\,\frac{1}{1+\ln\left(\frac{A}{2\epsilon_{k}}\right)}\epsilon_{k}^{2}+\ldots,\label{eq:pc2order}
\end{eqnarray}
where the relation in Eq.~(\ref{eq:epsilonk}), $A\left(1+\epsilon_{k}\right)^{-k}\sim2\epsilon_{k}$,
defines $\epsilon_{k}$. Unfortunately, we do not see any way to determine
the arbitrary constant $A$ in this expansion. We assume that it is
a reflection of the local nature of our expansion for $k\to\infty$
\citep{BO}. It seems that $A$ would have to be determined from global
information about the solution, such as from an evaluation at $n=0$.

\bibliographystyle{apsrev}
\bibliography{/Users/stb/Boettcher}

\end{document}